# Development Framework for Longitudinal Automated Driving Functions with Off-board Information Integration

Eric Armengaud, Sebastian Frager, Stephen Jones, Alexander Massoner, Alejandro Ferreira Parrilla, Niklas Wikström, Georg Macher

AVL List GmbH
Graz, Austria

Increasingly sophisticated function development is taking place with the aim of developing efficient, safe and increasingly Automated Driving Functions (ADF). This development is possible with the use of diverse data from sources such as Navigation Systems, eHorizon, on-board sensor data, Vehicle-to-Infrastructure (V2I) and Vehicle-to-Vehicle (V2V) communication. Increasing challenges arise with the dependency on large amounts of real-time data coming from off-board sources. At the core of addressing these challenges lies the integration of the Digital Dependability Identity (DDI) concept for reliable integration of off-board data. DDIs are modular, composable, and executable components in the field, facilitating:
- efficient synthesis of component and system dependability information,
- effective evaluation of information for safe and secure composition of highly distributed and autonomous Cyber Physical Systems.

In AVL's Connected Powertrain™, Automated Driving Functions are tailored to Powertrain Control Strategies that predictively increase energy efficiency according to the powertrain type and its component efficiencies. Simultaneously, the burden on the driver is reduced by optimizing the vehicle velocity, whilst minimizing any journey time penalty.

In this work, the development of dependable Automated Driving Functions is exemplified by the Traffic Light Assistant (TLA), an adaptive strategy that utilizes predictions of preceding traffic, upcoming road curvature, inclination, speed limits, and especially traffic light signal phase and timing information to increase the energy efficiency in an urban traffic environment. A key aspect of this development is the possibility for seamless and simultaneous development; from office simulation to human-in-the-loop and to real-time tests that include vehicle and powertrain hardware. Test results are shown from a simulated use case where a TLA-equipped ego vehicle is approaching an occupied pedestrian crossing. The use case demonstrates how cooperation with a preceding vehicle enables smoother and more energy-efficient maneuvers. The tests were carried out in a co-simulation toolchain which facilitates the transition to subsequent phases in the development cycle.

## 1 Introduction

In many domains, such as the automotive domain, a trend towards more collaboration of rather capsulated systems has become popular in the recent years. Before the introduction of wireless connections and automated driving functionalities, vehicles were physically isolated machines with mechanical controls. The emergence of cyber-physical automotive systems over the last decades has affected the development of vehicles, through the availability of external information (e.g. traffic information and connectivity features) promising ways to support new applications and altering the customer added value of the passenger car have been opened up.

In the context of the rising vehicle-to-vehicle and vehicle-to-infrastructure connectivity, which causes multiple inter-vehicle connections as well as capabilities for (wireless) networking with other vehicles and non-vehicle entities, automotive systems are developing from stand-alone systems towards systems of systems, interacting and coordinating with each other and influencing vehicle actions. Connections are not restricted to internal systems (e.g. steering, sensor, actuator, and communications) but also include other road users and the infrastructure. Future vehicles will be able to utilize connectivity features for over-the-air updates, integration of cloud-services, or remote automation.

The systems are integrated - more often dynamically at run-time - into so-called Systems of Systems (SoS), which consist of different collaborating systems (i.e. entities encompassing both hardware and software) that host several applications. Technologically, this is driven by ever more closely interconnected distributed embedded systems of systems running under the umbrella terms of cyber-physical systems (CPS) and Internet of Things (IoT).

In the course of this work, the development of dependable Automated Driving Functions (ADF) is exemplified by the Traffic Light Assistant (TLA) functionality. The TLA, an adaptive strategy that utilizes predictions of preceding traffic, upcoming road curvature, inclination, speed limits, and especially traffic light signal phase and timing information to increase the energy efficiency in an urban traffic environment. A key aspect of this development is the possibility for seamless and simultaneous development; from office simulation to human-in-the-loop and to real-time tests that include vehicle and powertrain hardware. This development requires a flexible toolchain which can facilitate transitions between the different development cycles, e.g. by utilizing modular interfaces which allows for simulation models to be easily exchanged by real hardware. In a first step, a simulated use case is presented in which a TLA-equipped ego vehicle approaches an occupied pedestrian crossing. The TLA selects velocities based on predictions made using currently available information, which can be enhanced using V2V communication.

The core concept of a Digital Dependability Identity (DDI) of a component or system is used to integrate off-board information into the decision-making of the TLA. This DDI is a modular, composable, and executable component contract, which facilitates the efficient synthesis of component and system dependability information, evaluation of information for safe and secure composition of highly distributed and autonomous Cyber Physical Systems at runtime. Driver's acceptance and comfort will in the future be rated in an advanced diver simulator mounted on a hexapod, capable of emulating longitudinal and lateral acceleration of a real vehicle.

The paper is organized as follows: Section 2 presents an overview of the relevant technology bricks of this work. In Section 3 a description of the use case in which the presented technology bricks are integrated is presented. Finally, Section 4 concludes with an overview of the approach presented.

## 2 Technology Bricks

In this section an overview of the relevant technology bricks of this work are given. AVL's Connected Powertrain™ is a library of automated driving functions and powertrain control strategies that is developed to increase energy efficiency of the powertrain type and its components.

A further focus is set on the seamless and simultaneous development framework, from office simulation to human-in-the-loop and to real-time tests that include vehicle and powertrain hardware.

The third technology brick briefly describes the concept of a Digital Dependability Identity (DDI); a modular, composable, and executable component contract.

### 2.1 Predictive Energy Management linked to Automated Driving Functions

The main objective of an Automated Driving Function (ADF) is the automatic choice of the velocity and steering wheel angle for any given driving situation. For powertrain energy management strategies only the velocity plays a key role, whereas the steering wheel angle can mostly be neglected in terms of energy efficiency. Autonomous driving functions have access to detailed information about the vehicle surrounding (e.g. road, speed limits, traffic etc.) from merging off-board and on-board data. For example, sensor data is used to capture motion profiles of other traffic participants. Additionally, information about the road topology, legal speed limits, curvature or traffic light signal phasing amongst other aspects are also considered ADF. Due to the comprehensive awareness about the road characteristics and driving situation, autonomous driving functions typically can influence the future choice of velocity and therefore have some knowledge about the expected upcoming vehicle speed.

This information is especially used for predictive and energy efficient operation strategies of powertrains. AVL's Connected Powertrain uses the connectivity and information from automated driving functions to control battery management and other systems or components of the powertrain. This may be achieved by sensible selection of drive modes (e.g. electric driving or use of combustion engine), tailored to current and upcoming traffic conditions, road gradient, curvature and other factors. Thus, on one hand predictive energy management strategies benefit greatly from automated driving functions. On the other hand, automated driving functions such as the traffic light assistant (TLA) can include detailed knowledge of the powertrain topology, component characteristics and states which are considered in the computation of energy-efficient velocity profiles, whilst minimizing journey penalty time. Further details on the TLA are described below.

## 2.2 Traffic Light Assist (TLA) for Automated Driving

One recurring every-day scenario in traffic is the need for a vehicle to stop at a precisely defined location at a precisely defined time. Such events occur for example when approaching a red traffic light. Increasingly intelligent vehicles, however, may in many cases use a redefined approach, not needing to perform a complete vehicle stop. Instead of stopping the vehicle completely, a requirement to not overstep a precise location before a specific point in time often allows more efficient and safe handling of various driving situations. Although some human drivers with an anticipatory driving style may follow a somewhat similar approach, an intelligent vehicle equipped with automated driving functions, sensor fusion and c2x communication installed, typically can gather and process far more data (especially with regard to information outside line of sight) about the surrounding than a human driver would be able to.

The on-board Traffic Light Assistant (TLA) system [1] uses an optimization algorithm considering the vehicle and powertrain characteristics as well as the traffic conditions ahead of the vehicle. Model Predictive Control (MPC) is employed to obtain optimal speed profiles to deal with multiple upcoming road junctions equipped with Traffic Lights. MPC is a finite-time optimal control methodology that considers future system behavior using a prediction model and minimizes a defined system cost. As the calculation of the cost also includes detailed powertrain component efficiencies and limitations (e.g. the maximum regenerative braking power of a hybrid vehicle), the resulting velocity trajectory is tailored to a specific vehicle in a specific situation.

MPC computes the optimal sequence of the control inputs in a receding-horizon approach. This means that even though an optimal control sequence is computed for a defined control horizon, only the optimal control inputs corresponding to the actual time are applied to the system. Therefore, the optimization routine has to be repeated for every time step to obtain the optimal control input. However, the rest of the control sequence $u_{opt,prev}$ is used for the prediction model. In this sense, the model represented by state-space equation is expanded along the control horizon and prediction horizon such that,

$$xpred = M * x(k) + N * uopt, prev$$
$$ypred = E * x(k) + F * uopt, prev$$

The matrices $M$, $N$, $E$ and $F$ are obtained recursively from the system definition of a state space model that represents vehicle and powertrain. The variable $x(k)$ describes the system states at time k, whereas $x_{pred}$ and $y_{pred}$ refer to the predicted system states and system outputs, respectively.

The region of the optimal solution (i.e. feasible intervals for the control sequence) is either constrained by the dynamical model of the system, in this case, longitudinal vehicle and powertrain related constraints, or by the external conditions such as traffic light position, timing of green or red phases or other vehicle interactions. MPC includes a model based prediction approach and minimizes a defined cost function considering various state, input and output constraints every sampling interval. The cost function quantifies the energy consumption, comfort and travel time within the prediction horizon:

$$min \sum_{\tau=t}^{t+N_p} f(x(\tau), u(\tau))$$
$$\text{S.T. } g(x(\tau), u(\tau), \tau) \leq 0$$
$$u(\tau) \in U, \quad x(\tau) \in X, \quad \tau = 1, \dots, N_p - 1$$
$$\tau = t, \dots, t + N_p - 1$$
$$x: state\ variables, u: control\ variables, \tau: time,$$
$$N_p: prediction\ horizon$$

The sum of the objective function $f(x(\tau), u(\tau))$ is minimized over the prediction horizon $N_p$, subject to the system constraints $g(x(\tau), u(\tau), \tau)$ which limit possible future states of the vehicle, e.g. (position, velocity) to be within certain constrained boundaries. As seen in Figure 1, a red traffic light signal phase represents a time-dependent position constraint which the vehicle must respect until the signal phase turns green. Briefly, a vehicle with a certain initial velocity, $v_0$, aims to be on position $x_{TL1}$, $x_{TL2}$...etc. at certain time $t$ specified by phases of the traffic lights. These phases are characterized by $[t_{gi} \leq t \leq t_{ri}]$ where $t_{ri}$ specifies green to red switching time and $t_{gi}$ red to green switching time:

$$pos \in R_I \quad \text{if} \quad 0 \leq t < t_{g1}$$
$$pos \in R_{II} \quad \text{if} \quad t_{g1} \leq t < t_{r1}$$
$$pos \in R_{III} \quad \text{if} \quad t_{r1} \leq t < t_{g2}$$
$$pos \in R_{IV} \quad \text{if} \quad t_{g2} \leq t < t_{r2}$$
$$\dots$$

$$v \geq 0$$

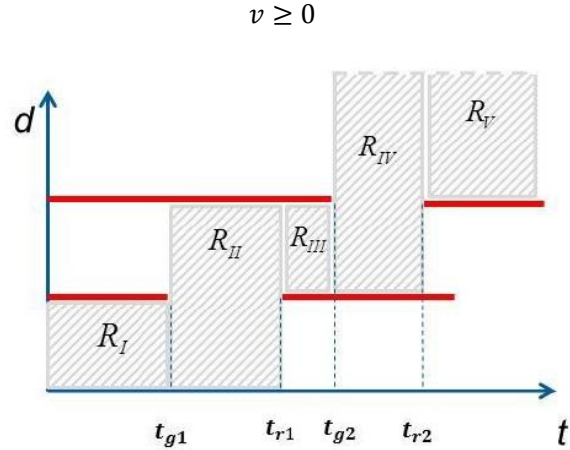

Figure 1: Red traffic light signal phases imposing time-dependent distance constraints.

An important factor for the accuracy and optimality of the prediction is the confidence of traffic light signal phase information. It is known that predicting the duration of red/green phases for a long time horizon is a difficult task, since such traffic light programs depend on various environmental factors, including time of day and traffic load [2]. Meaningful ancillary information of broadcasted traffic light signals via V2X communication therefore includes some indication of the confidence or probability associated with the transmitted data. Using this information, a probability density function for points in time, at which traffic lights ahead may switch to green can be estimated. This allows useful modifications to various constraints of the optimization problem, subject to the uncertainty over the prediction horizon. Modifications may include for example travel time constraints related to traffic light signal phase timing information.

The method described above is in principal applicable also to compute optimal control sequences when approaching pedestrian crossings, since pedestrians can be interpreted in a similar way to traffic lights by the formulation of the optimization problem. It is assumed that an on-board camera is able to identify pedestrians or other relevant objects in the vicinity of pedestrian crossings. This information is merged with available information about traffic light signal phase timing obtained via camera or V2X communication as described above. In Figure 2 it is shown that the vehicle is not allowed to trespass a certain position at certain times (i.e. while the pedestrian is crossing the road). If a pedestrian is located at the edge of the road and starts crossing the road at a certain velocity $v_p$, the time that the pedestrian will need to safely cross the road can be estimated. The location and time it takes the pedestrian to cross the street represents constraints implemented as time dependent position constraints in a similar manner to the traffic light constraints. Of course, safety has to be considered at all moments, and the position of the pedestrian and estimated trajectory must be monitored constantly to ensure safety. Briefly, consider a pedestrian crossing located at a position $v_p$. At time $t_p$, a pedestrian starts crossing the road with an estimated constant velocity $v_p$. The estimated duration of the 'virtual red phase' $t_{red}$ is assumed to be sufficient to allow a safe pedestrian crossing.

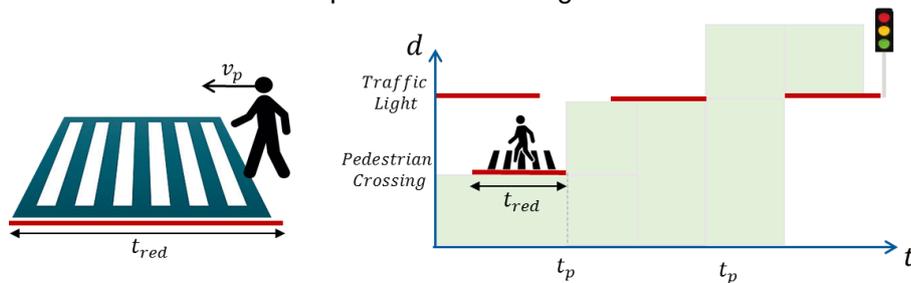

Figure 2: Crossing pedestrians imposing time-dependent distance constraints

Note that the presented approach is not the only way to formulate pedestrians as constraints; a pedestrian can alternatively be modelled as a position dependent velocity constraint set to zero (i.e. the vehicle has to stop if it reaches the position of the crossing pedestrian). The constraints imposed by Traffic lights as well as the constraints imposed by pedestrians are then merged, and the minimum of both is selected at each time instant within the prediction horizon.
Additionally, a travel time reference can be introduced as soft constraint to the optimization problem. The target is to increase road throughput by encouraging vehicles to pass traffic lights as soon as they switch to green. To penalize deviations from this behavior, a travel time reference can be constructed as seen in Figure 3. The basis of the reference is calculated by assuming constant velocity between two traffic lights and reaching both traffic lights at the earliest feasible point in time when they switch to green. The first part of the reference is constructed by calculating the speed needed to exactly reach the next green phase from the current vehicle position. Starting from the ego vehicle's position, the speed needed to exactly reach the next green phase is calculated. If this speed exceeds the maximum allowed legal fixed speed limit, the speed is saturated to the legal speed limit. The resulting speed is then used to calculate a corresponding position trajectory reference over the prediction horizon. An additional cost (penalty) is given to any solutions which would be slower than the reference trajectory at any point in time. The derived reference trajectory can alternatively be replaced by a terminal cost associated with the desired position at the end of the prediction horizon ($d_{Hp}$).

Note that this reference trajectory can be extended to the case where a preceding vehicle is detected in such a way that the reference position is saturated with the preceding vehicle's predicted position, and a safety distance is always guaranteed. Furthermore, if multiple sources of information are available with their associated confidence levels, e.g. concerning the starting times of green phases or the speeds of preceding vehicles, the reference trajectory can be softened or tightened to account for the varied uncertainty over the prediction horizon using stochastic modelling techniques such as individual chance constraints [3].

Please note that although not explicitly described, system constraints include other traffic participants, legal speed limits, road properties & curves to the optimization strategy.

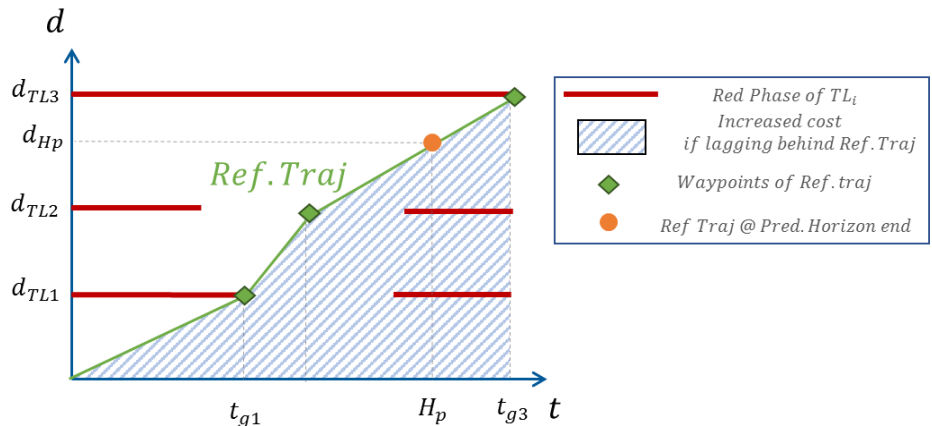

*Figure 3: Position reference trajectory & increased cost region*

## 2.3 Development Environment / Toolchain

Providing the highly sophisticated functionality of the TLA, the optimization algorithm has to be supplied with data from the environment, such as upcoming traffic, curvature or traffic light phases, the current powertrain status and data from the current state of the vehicle. The environment, powertrain and vehicle simulation models are interacting with each other and also with the TLA algorithm. Testing and validation of the exchanged data between these systems is essential to guarantee a failsafe and working algorithm, therefore AVL Model.CONNECT™ is used as a simulation platform for the co-simulation and the data exchange of all simulation models. This tool offers the possibility to connect different AVL or third party tools and build up a co-simulation model. Coupling models with different solver time steps demands inter- or extrapolation of exchanged data. This can lead to coupling errors and time delays. AVL Model.CONNECT™ offers a self-developed coupling method, the so called Nearly Energy Preserving Coupling Element(NEPCE), which minimizes this coupling errors and improves the overall stability of the system. The complete vehicle simulation is distributed between two further AVL tools. The powertrain, including hybrid components, combustion engine, drive shafts up to the wheel hubs is represented by AVL CRUISE™. The vehicle body, including chassis, steering system, tires is simulated by AVL VSM™. This precise description of the overall vehicle behavior, consisting of the complete hybrid powertrain and the chassis, is used for the model predictive controller. The vehicle interacts and receives information from an environment simulation model, for this purpose Vires Virtual Test Drive (VTD) is used. It offers the possibility to set up an environment consisting of traffic, road networks, traffic signs and lights and it can provide this data to other tools, for example for Vehicle-to-Vehicle or Vehicle-to-Infrastructure communication. The possibility to control each traffic light independently offers a validation opportunity. Corrupted data can be sent to test the stability and robustness of the TLA and different light phases can be used depending on the current time of the day. The last component integrated in the co-simulation is the Traffic Light Assist. Depending on the software basis and the programming code it is integrated and connected with the environment, powertrain, and vehicle model. For the integration of the TLA as a Simulink model into the co-simulation platform the Independent Co-Simulation (ICOS) system is used, which offers the possibility to exchange data between AVL Model.CONNECT™ and a Simulink model.

The first step is to set up a running and stable co-simulation of all tools in an office environment to validate the exchanged data. This data has to be plausible and exchanged in the correct data format in order to guarantee a failsafe system which works correctly. In this office environment a first estimation of the overall system can be done and different simulation scenarios can be set up. One scenario could be to introduce a difference between the information coming from the traffic light and the delivered information from the car's sensors. The traffic light sends information that the green light phase is currently active but the car's camera sends the information that the traffic light shows a red light. This can be used to test and evaluate the decision making algorithm and to test which information can be trusted more.

With a stable and running office simulation model it is possible to connect with the advanced driver simulator and the hexapod to test driver's acceptance and comfort. The vehicle's longitudinal, lateral and vertical accelerations are sent to a hexapod which translates this information into motion. This configuration offers the future possibility to subjectively evaluate the TLA in regard to comfort and hazard perception, e.g. distance to vehicle ahead or velocity profile during traffic light approach. AVL Model.CONNECT™ offers an Advanced Co-Simulation for Realtime Applications (ACoRTA) coupling method to connect office simulation models with hardware components and run the system in real-time. By connecting the simulation model with AVL-DRIVE™ the subjective evaluation can also be compared with an objective evaluation. Regarding driver comfort, AVL VSM™ can simulate an elastically mounted powertrain. With this, the TLA can be optimized regarding powertrain vibrations which are transferred to the driver and can lead to dizziness and discomfort.

## 2.4 The Digital Dependability Identity (DDI) Concept

In general, a Digital Identity is defined as "the data that uniquely describes a person or a thing and contains information about the subject's relationships" [4]. Applying this idea, a DDI contains all the information that uniquely describes the dependability characteristics of a system or component. This includes attributes that describe the system's or component's dependability behavior, such as fault propagations, as well as requirements on how the component interacts with other entities in a dependable way and also the level of trust and assurance that can be guaranteed. The meta-information of the DDI are produced during design and continually maintained over the complete lifetime of a component.

Essential for the realization of the DDI concept are: (a) an open meta-model for specification of DDIs and exchange of dependability information, and (b) a concept for the dynamic integration and evaluation of DDIs. The DDI approach uses a language subset, which was created to describe the contracts based on an extension of XML. The language ties together basic meta-information of demands and guarantees and facilitates the automated evaluation process. The demands and guarantees are currently focusing on safety requirements. Safety requirements are either requested to be fulfilled by somebody (demand) or are guaranteed to be fulfilled (guarantee). They always consist of a statement (i.e., what is guaranteed or demanded) together with a level of confidence with respect to the actual fulfillment of that statement, and are currently used for binary decisions whether or not to accept the specific item. More details on the DDI concept are given in [5].

DDIs are used for the integration of components to systems during development as well as for the dynamic integration of systems to "systems of systems" in the field. To achieve this vision, we develop DDIs as modular, composable, and executable model-based assurance cases with interfaces expressed in SACM and internal logic that may also include fault trees, state automata, Bayesian networks and fuzzy models for representing uncertainty. Modularity means DDIs apply to units at different levels of design, including the system itself, its subsystems and components. By being composable, the DDI of a unit can be derived in part from its constituent elements.

The DDI approach application for the TLA system allows integrating also off-board information sources for the ego-vehicle to follow the optimal velocity trajectory to efficiently drive a 'Green Wave'. For this purpose, the TLA relies on V2I communication via Wi-Fi. This V2I communication makes use of the DDI to decide whether or not to accept the specific service information (traffic light green and red phase and position information) from the traffic light or make use of as well as possible pedestrian recognition by other vehicles. Figure 4 depicts an excerpt of the DDIs for a cloud-based corner steering service, which includes also the TLA functionality.

The DDI approach is another technology brick, which is applied in the development framework context of this work. More details regarding the DDI application in the automotive context was subject to publication [6] and is not included in this work.

```
 1 <DDI>
 2 <ComponentName> Cloud-based Corner Steering Service
        </ComponentName>
 3 <Guarantee>
 4 <ConfigurationName> CornerSteering </ConfigurationName>
 5 <IntegrityLevel> D </IntegrityLevel>
 6 <SecurityProperty> 3 </SecurityProperty>
 7 <DemandSet>
 8   <Demand>
 9     <ConfigurationName> acceleration </ConfigurationName>
10     <IntegrityLevel> D </IntegrityLevel>
11   </Demand>
12   <Demand>
13     <ConfigurationName> Lane Keep Assistant </ConfigurationName>
14     <IntegrityLevel> D </IntegrityLevel>
15   </Demand>
16   <Demand>
17     <ConfigurationName> emSpeed </ConfigurationName>
18     <IntegrityLevel> B </IntegrityLevel>
19   </Demand>
20   <Demand>
21     <Platform_Service>
22     <Failure> Lane Keep Assistant Failure </Failure>
23     <Reaction> detected </Reaction>
24     <IntegrityLevel> D </IntegrityLevel>
25     <Error> 3 % </Error>
26     </Platform_Service>
27   </Demand>
28   <Demand>
29     <HealthMonitoring>
30     <Failure>
31     <Application> Application Runtime Failure </Application>
32     <ApplicationResourceName> Lane Keep Assistant
         </ApplicationResourceName>
33     <Latency> more than 10 ms </Latency>
34     </Failure>
35     <IntegrityLevel> D </IntegrityLevel>
36     </HealthMonitoring>
37   </Demand>
38 </DemandSet>
39 </Guarantee>
40 </DDI>
```

*Figure 4 Excerpt of DDI for Cloud-based Corner Steering Service*

## 3    Use Case Description

The use case introduces a co-simulation toolchain based on AVL VSM[1] for the comprehensive evaluation of complex autonomous driving functions. The proposed environment contains an extensive package of tools and services that support the OEM in the prediction of vehicle behavior, and enables improvement of various vehicle attributes from the initial concept to the testing phase.
An approach for the optimal control of a fully electric vehicle and its powertrain approaching a road segment with Multiple Traffic Lights (TL) has been presented in [1] and briefly introduced in Section 2. A system referred to as the Traffic Light Assistant (TLA) was developed in order to take over longitudinal control of the vehicle, optimizing the velocity trajectory when approaching multiple traffic lights in traffic (see Section 2.2). The main aims of the system are to reduce energy consumption and $CO_2$ emission, reduce the number of vehicle stops and waiting times, and allow for smoother speed profiles (see Fig. 3). The presented TLA approach has since been further developed to work with other powertrain topologies [7], [8] and was demonstrated in real-time for conventional vehicles on an AVL powertrain testbed within the FFG Austrian Funded R&D Project TASTE (Traffic Assistant Simulation and Testing Environment).

---

[1] https://www.avl.com/-/avl-vsm-vehicle-simulation

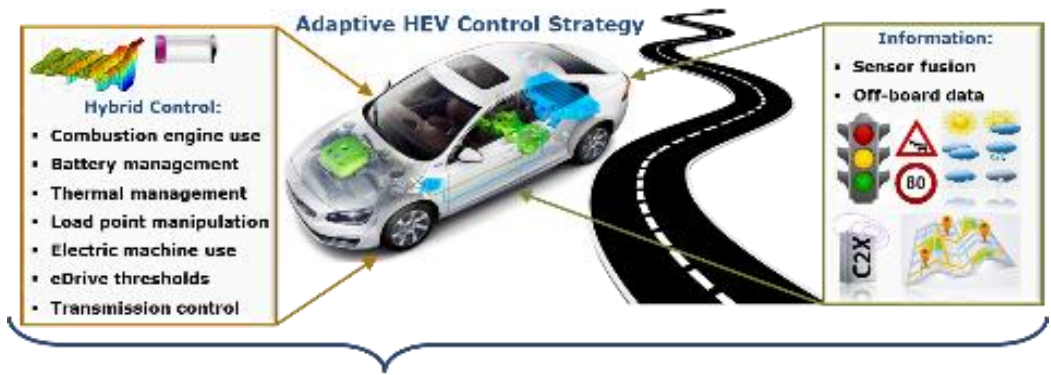

*Figure 5 Connected Powertrain and Autonomous Driving Functions Use Case*

Previous work has focused on controlling an ego vehicle equipped with the TLA, assuming complete knowledge about the road conditions and traffic light signal phasing. In reality, not all information may be accessible at all times to every vehicle. For example, if the current state of a traffic light or a pedestrian at the side of a crossing is observed by an on-board camera or local V2I communication, the information would only be available to a vehicle close by. Such a vehicle, if equipped with automated driving functions, would adapt its velocity accordingly to stop in front of the traffic light, or to safely let the pedestrian pass. However, a following vehicle might not directly see the traffic light or the pedestrian and therefore could not anticipate the behavior of the preceding vehicle. In this case, a reaction can only occur based on observation of the behavior of the preceding vehicle. It is possible to improve energy efficiency, if the state of traffic lights or the presence of pedestrian are known in advance. Once more the opportunity for increased efficiency and safety comes with increased security threats. To that aim, DDIs will explore how to improve control strategies to increase energy efficiency while ensuring safety at the same time. In the depicted DDI concept, the infrastructure provides services for the connected car (such as cornering speed recommendations for the specific vehicle type) plus an additional service DDI. Based on the information provided by the DDI, the vehicle can generate and evaluate the dependability of the specific configuration with the additional service provided by the infrastructure and can decide whether or not to make use of the (temporally) available service and adapt to the current situation.

The DDI approach is applied for a traffic light assistant (TLA) system, which allows ego-vehicles to follow optimally calculated (on-board vehicle) velocity trajectory to efficiently drive 'Green Wave', see Section 2.2. For this purpose, the TLA relies on V2I communication, via Wi-Fi and also considers powertrain dynamic states. This V2I communication makes use of the DDI to decide whether or not to accept the specific service information (TL green and red phase and TL position information) from the traffic light or make use only of on-board systems.

The further developed Traffic Light Assistant is applied on the Use Case shown in Figure 6 where 2 vehicles approach a crosswalk and a pedestrian is waiting to safely cross the road. Vehicle (1) will detect the pedestrian before vehicle (2). Passing on this information to vehicle (2) allows vehicle (2) to react earlier and in a more energy-efficient way. Vehicle (1) also estimates the time it will take the pedestrian to cross the road, similar to the time until a red traffic light switches to green. Alternatively, vehicle (1) could directly advise a recommended velocity for vehicle (2). Vehicle (2) could then follow the recommendation as long as specific criteria are met, ensuring safe, reliable and comfortable operation.

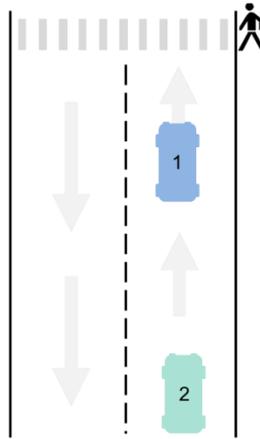

*Figure 6 Two vehicles approaching a cross-walk with waiting pedestrian use case*

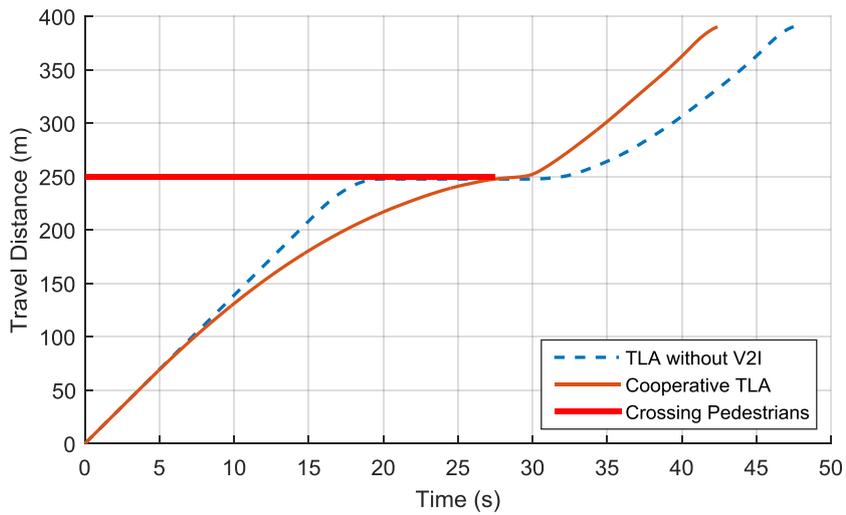

*Figure 7 TLA travel distance over time when approaching pedestrian crossing.*

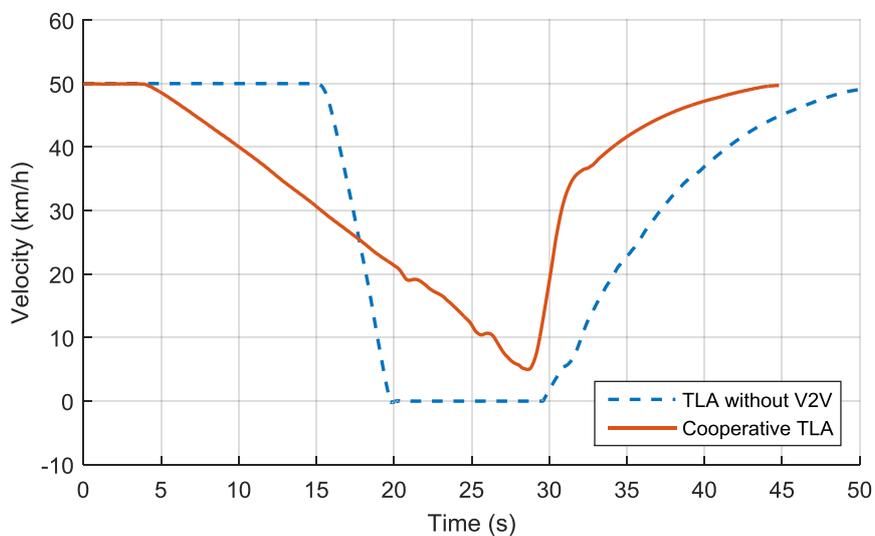

*Figure 8 TLA speed profile when approaching pedestrian crossing.*

Figure 7 and Figure 8 show the TLA-equipped ego vehicle's speed profiles when approaching a pedestrian crossing located 250 meters ahead, travelling at an initial velocity of 50 km/h. In the first case, the ego vehicle only has access to information from on-board sensors, e.g. a camera-based pedestrian detection system, with a maximum detection range of 50 meters. In the second case, the ego vehicle cooperates with the preceding vehicle, which is located at the pedestrian crossing and broadcasts that it is being occupied to the ego vehicle. The V2V broadcast range was in this case assumed to be 200 meters. The cooperative variant of the TLA performs a smoother approach to the crossing, avoiding a complete stop. This allows the cooperative TLA to consume 19.6% less energy for the complete maneuver, which in both cases ends at 150 meters after the pedestrian crossing.

## 4  Conclusion

In this work, the development of dependable automated driving functions is exemplified by the traffic light assistant, an adaptive strategy that utilizes predictions of preceding traffic, upcoming road curvature, inclination, speed limits, and especially traffic light signal phase and timing information to increase the energy efficiency in an urban traffic environment. A key aspect of this work is the possibility for seamless and simultaneous development; from office simulation to human-in-the-loop and to real-time tests that include vehicle and powertrain hardware.

For that aim, an overview of the relevant technology bricks: AVL's Connected Powertrain™ library of automated driving functions and powertrain control strategies, the seamless and simultaneous development framework, and the concept of a digital dependability identity (DDI); a modular, composable, and executable component contract is given. The concept was demonstrated on a simple V2V-based use case, where a Traffic Light Assistant cooperates with a preceding vehicle while approaching a pedestrian crossing, providing a smoother transition with a significant energy consumption improvement.

## 5  Acknowledgments

This work is supported by the DEIS project – Dependability Engineering Innovation for automotive CPS. This project has received funding from the European Unions Horizon 2020 research and innovation programme under grant agreement No 732242.